\documentclass[aps,pra,twocolumn,showkeys,showpacs]{revtex4}
\usepackage{graphicx}

\def\ket#1{|\,#1\,\rangle}
\def\bra#1{\langle\, #1\,|}

\def\opone{\leavevmode\hbox{\small1\kern-3.8pt\normalsize1}}

\newcommand{\beq}{\begin{equation}}
\newcommand{\eeq}{\end{equation}}
\newcommand{\be}{\begin{eqnarray}}
\newcommand{\ee}{\end{eqnarray}}
\newcommand{\bea}{\begin{eqnarray}}
\newcommand{\eea}{\end{eqnarray}}
\newcommand{\bma}{\begin{subequations}}
\newcommand{\ema}{\end{subequations}}
\newcommand{\bwt}{\begin{widetext}}
\newcommand{\ewt}{\end{widetext}}

\newcommand{\YSO}{Y$_2$SiO$_5$}

\begin{document}
\title{Photon-Echo Quantum Memory}
\author{W. Tittel}
\email[]{wtittel@qis.ucalgary.ca}
\affiliation{Institute for Quantum Information Science \& Department of Physics and Astronomy,
University of Calgary, Canada}

\author{M. Afzelius}
\affiliation{Group of Applied Physics, University of Geneva, Switzerland}

\author{R.L. Cone}
\affiliation{Department of Physics, Montana State University, Bozeman, USA}

\author{T. Chaneli\`{e}re}
\affiliation{Laboratoire Aim\'e Cotton, CNRS-UPR 3321, Orsay, France}

\author{S. Kr{\"o}ll}
\affiliation{Department of Physics, Lund University, Sweden}

\author{S.A. Moiseev}
 \affiliation{Kazan Physical-Technical Institute of the Russian Academy of Sciences, Russia
 \footnote{current address: Institute for Quantum Information Science \& Department of Physics and Astronomy,
University of Calgary, Canada}}

\author{M. Sellars}
\affiliation{Laser Physics Centre, Australian National University, Canberra, Australia}

\begin{abstract}
The future of long-distance quantum communication relies on the availability of quantum memory,
i.e. devices that allow temporal storage of quantum information. We review research related
to quantum state storage based on a photon-echo approach in rare earth ion doped crystals and glasses.
\end{abstract}

\keywords{quantum memory, quantum repeater, quantum communication, photon-echo, rare-earth-ions}

\pacs{01.30.Rr, 03.67.Hk, 42.50.Ex, 78.47.jf, 32.80.Qk, 42.70.Ln}

\maketitle

\section{Introduction}
\label{Introduction}

Optical data storage has been an important research topic for many years (see, e.g., \cite{all_optical_storage}). It emerged and grew with the development of information and communication technology, in particular with the demand for fast data access and storage capacity. The development of quantum information theory and quantum communication \cite{Nielsen2000,PhysicsWorld}, which promises unprecedented computational capacities via quantum computing and unconditional communication security via quantum cryptography, has recently added additional interest to some approaches to optical storage. As is the case in classical information processing, quantum information processing also requires temporal storage for (quantum) data. However, a simple classical 'measure and write down' approach is impossible for storage of quantum information. Indeed, the measurement process would change the encoded quantum information in an irreversible way. This cornerstone of quantum information theory is deeply rooted in the foundations of quantum theory. It emerges from the well-known Heisenberg's uncertainty relations and is generally referred to as the no-cloning theorem (for a recent review see \cite{Scarani2005}).

The challenge of reversibly transferring quantum information between photons, i.e. moving carriers suitable for sending quantum information, and atoms, i.e. stationary carriers for storage, has recently triggered the development of quantum state storage protocols.
In this article, we will review work related to an approach based on controlled reversible inhomogeneous broadening of a single atomic absorption line (CRIB)\cite{Moiseev2001,Nilsson2005a,Alexander2006,Kraus2006} (for other approaches to a light-matter quantum interface see \cite{Cirac1997,Fleischhauer2002,McKeever2002,Kuhn2002,Mundt2002,Julsgaard2004,Moehring2004,Chou2004,Chaneliere2005,Eisaman2005,Gorshkov2007}). It results from the observation that a pulse of light, absorbed in an inhomogeneously broadened medium with small homogeneous linewidth, can be forced to re-emerge from the medium some time later as an echo (see Fig. 3). The so-called photon-echo has been proposed  independently by Kopvil'em \textit{et al.} \cite{Kopvil'em1993} and Kurnit \textit{et al.} \cite{Kurnit1964} (who has also reported the first experimental demonstration) in 1963 and 1964, respectively,  as the optical analog to the well-known spin echo, discovered by Hahn in 1950 \cite{Hahn1950}. In the late seventies and early eighties, Elyutin and Mossberg then independently extended this idea and proposed a way to use three pulse photon-echoes for storage of classical data \cite{Elyutin1979, Mossberg1982}. While impressive results have meanwhile been obtained, including storage and recall of a data sequence consisting of 1760 optical pulses \cite{Lin1995}, traditional photon-echo based storage generally suffers from strong limitations when used for storage of single photons, and can thus not be directly exploited for quantum communication. Yet, it gave rise to quantum state storage based on CRIB, which is well suited for storage of single photons carying quantum information.

This article is organized in the following way: in section \ref{why QM} we will first introduce some basic notions and tools of quantum communication, and then briefly elaborate on the role of quantum memory for future long distance quantum communication links as an important part of a quantum repeater. Section \ref{History&CRIB} reviews the historical development from photon-echoes to CRIB-based quantum memory, and gives a simple theoretical description of the quantum memory protocol. This section is followed by a discussion of material properties of rare-earth-ion (RE) doped crystals and glasses in view of CRIB and quantum repeaters.
Section \ref{photon echoes as test bed} summarizes recent experiments that study traditional photon-echo based data storage for its suitability to store and recall amplitude and phase information encoded into subsequent pulses of light, which is a necessary condition for storage of quantum information. This section is followed by a short presentation of experimental demonstrations of the new quantum memory protocol for storage of strong pulses (section \ref{experimental CRIB}). The article finishes with a conclusion and outlook.

\section{Quantum memory for quantum repeater}
\label{why QM}
The possibility to store and recall non-classical light states on demand, including single photons or photons that belong to entangled pairs, would significantly benefit many applications of quantum information processing. It would allow building a triggered single-photon source based on a heralded but probabilistic source \cite{Matsukevich2006} and would thus remove a potential security hole in quantum cryptography based on faint laser pulses \cite{Brassard2000} (this threat can also be removed by resorting to quantum cryptography protocols employing entanglement \cite{Ekert1991}, or decoy states \cite{Hwang2003,Wang2005,Lo2005}). Furthermore, quantum memory is a key ingredient in linear optics quantum computing (see \cite{Kok2007} for a recent review). In this article, we are mainly interested in the role of quantum memories for long distance quantum communication, more precisely for quantum repeaters. In this section, we will first very briefly discuss some basic notions of quantum communication (more complete presentations on various aspects can be found in the book by Nielsen and Chuang\cite{Nielsen2000}, and in review articles by Tittel and Weihs \cite{Tittel2001}, Gisin \textit{et al.}\cite{Gisin2002}, and Pan \textit{et al.}\cite{Pan2008}), and then discuss quantum memory in the context a quantum repeater, which was introduced by Briegel \textit{et al.} in 1998 \cite{Briegel1998}.

\subsection{Some quantum communication tools}
\label{Tools}

Quantum communication relies on exchanging quantum information encoded into quantum states between two (or more) parties, usually called Alice and Bob. In this article, we will restrict ourselves to the most frequently used approach to quantum communication, which employs quantum bits (or qubits), and we will limit our examples in this section to quantum information encoded into photons.

A qubit is generally described by
\beq
\label{qubit}
\ket{\psi}=\alpha\ket{0}+\beta e^{i\phi}\ket{1}
\eeq
where the orthogonal ket states $\ket{0}$ and $\ket{1}$ form a basis in an abstract, two dimensional Hilbert space, and $\alpha$, $\beta$ and $\phi$ are real parameters that determine the probability amplitudes and phase of these superposition states, respectively. Often, the qubit basis states are encoded into polarization states of photons, for instance right and left circular polarization states. Superpositions of the form of Eq. \ref{qubit} with equal probability amplitudes $\alpha$ and $\beta$ then include horizontal, vertical, and diagonal and anti-diagonal polarized photons. Another possibility to realize qubits is to use photonic wavepackets, which, at some given time $t$, are localized at positions $x_0$ and $x_1$, respectively -- so-called time-bin qubits. Hence, in this case, $\alpha^2$ and $\beta^2$ describe the probabilities for detecting a photon in the first, or second 'time-bin'.

The quantum mechanical superposition principle, which is at the heart of Eq. \ref{qubit}, can be generalized to multi-particle systems. For the case of two qubits A and B, this leads to states of the form $\ket{\psi}=\alpha\ket{a_1}_A\otimes\ket{b_1}_B+\beta e^{i\varphi}\ket{a_2}_A\otimes\ket{b_2}_B$ where the kets denote orthogonal basis vectors spanning the Hilbert spaces of A and B. Examples, as before, include polarization and time-bin states. Due to their peculiar non-local properties, entangled states have been subject to numerous fundamental theoretical and experimental investigations, which where triggered by the seminal papers by Einstein, Podolsky and Rosen in 1935 \cite{Einstein1935}, and Bell in 1964 \cite{Bell1964} (for reviews on entanglement see \cite{Tittel2001,Pan2008}). In addition, entangled states form the very key ingredient for quantum communication. Of particular interest are the four so-called Bell states

\bea
\label{Bell states}
\ket{\phi^{\pm}}&=&2^{-1/2}(\ket{00}\pm\ket{11})\nonumber\\
\ket{\psi^{\pm}}&=&2^{-1/2}(\ket{01}\pm\ket{10})
\eea
\noindent
where $\ket{ij}$ is a shorthand for the tensor product between $\ket{i}_A$ and $\ket{j}_B$. Note that these states describe qubit-pairs in pure states (tr($\rho^2$))=1), but that each individual qubit is in a maximally mixed state ($\rho_{i}=tr_{j}\rho_{ij}=\opone /2$). Here, $\rho_i$ and $\rho_{ij}$ denote single or two-qubit density matrixes, respectively, and $tr$ is the trace or partial trace operation. The four Bell-states form a basis for any two-qubit state. This is exploited in quantum teleportation \cite{Bennett1993} and entanglement swapping \cite{Zukowski1993}, as discussed in the remainder of this section.

Let us now assume that we have one pair of qubits in a known Bell-state, and a single qubit in an unknown state. We make a joint measurement on the single qubit and one qubit out of the entangled pair, i.e. we project the joint state onto the basis spanned by the four Bell-states. As discovered by Bennett and co-workers in 1993 \cite{Bennett1993}, each of the four possible results of the projection measurement is associated with a simple, deterministic unitary operation to be performed on the remaining qubit so that it is transformed into the quantum state initially encoded into the single qubit. In other words, the Bell-state measurement allows teleporting the unknown quantum state from a qubit onto another qubit. Starting in 1997, teleportation with photonic qubits has been demonstrated repeatedly in various laboratories  (see e.g. \cite{Bouwmeester1997,Marcikic2003,Riedmatten2004,Zhao2004,Houwelingen2006,Zhang2006,Zhang2006b}) and recently even outside the laboratory \cite{Ursin2004,Landry2007}.

Finally, let us consider the case where the qubit to be teleported is entangled with another qubit, i.e. we have initially two maximally entangled pairs. The joint measurement on two qubits from different pairs then leads to entanglement between the two remaining qubits in one of the four Bell states, determined, as in the case of teleportation, by the outcome of the entangling operation. This protocol has been discovered by \.{Z}ukowski and coworkers in 1993 \cite{Zukowski1993} and is now generally referred to as entanglement swapping or teleportation of entanglement. Experimental demonstrations with photonic qubits have been reported in \cite{Pan1998,Jennewein2001,Riedmatten2005,Yang2006,Halder2007}. Entanglement swapping constitutes a key ingredient in the quantum repeater, as discussed below.

\subsection{Quantum repeaters}
\label{Quantum repeater}
Most quantum communication protocols rely on close-to-perfect entanglement, that is Alice and Bob each possessing one photon out of a nearly maximally entangled pair in a state given by Eqs. \ref{Bell states}. For instance, in the case of quantum cryptography, Alice and Bob could perform single qubit measurements on their respective photons and thereby establish a secret key, as proposed by Ekert in 1991 \cite{Ekert1991}. Another use of entanglement, pointed out by Bennett and coworkers in 1993, is the faithful transfer of an unknown quantum state through quantum teleportation \cite{Bennett1993}. Unfortunately, the distribution of entanglement over long distances suffers from photon loss during transmission. For instance, assuming transmission of 1550 nm wavelength photons through an optical fibre with attenuation coefficient $\textsl{a}$=0.2 dB/km, the probability for successful entanglement distribution $P=10^{-\textsl{a} L/10}$ yields 0.1, 0.01, and $10^{-20}$ for distances $L$ of 50 km, 100 km, and 1000 km, respectively. Furthermore, the purity of the distributed entanglement decreases exponentially with the length of the quantum channel, due to detector noise or decoherence. Hence, quantum communication based on direct transmission of entanglement is limited to distances of the order of 100 km.

\begin{figure}[h]
  \includegraphics[width=.45\textwidth]{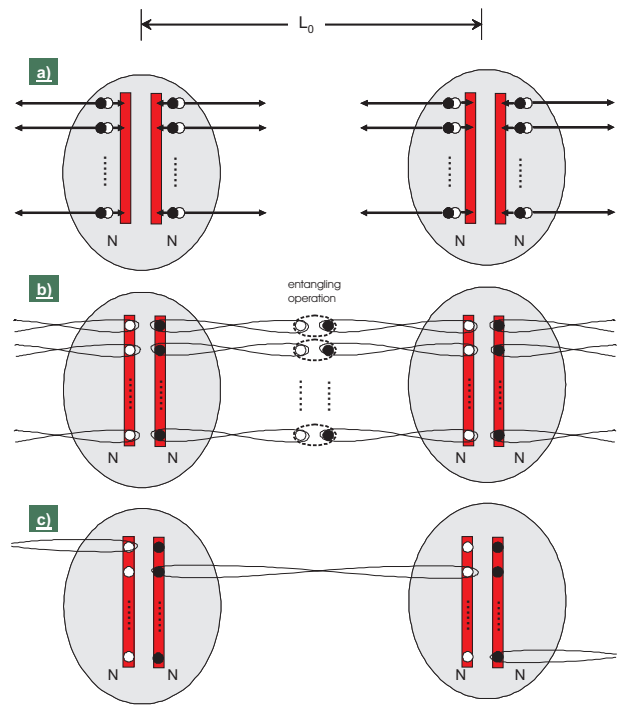}
  \caption{Schematical representation of two adjacent nodes in the simplified version of a quantum repeater. Each node (light grey circles) contains 2 N sources of entangled photon pairs (denoted by $\bullet\circ$), and two multi-mode quantum memories (denoted by red squares). One photon from each pair is stored in the quantum memory, and the other photon is sent towards one adjacent node (Figure a). Photons from different nodes meet half way between sources (provided they were not absorbed during transmission), where they are pairwise subjected to an entangling operation (Figure b). Entanglement is symbolized by the lying "figure of eight", the entangling operation (a Bell state measurement) by a dashed circle. The pairwise entanglement between photons and respective memory modes is thus swapped to entanglement between memories in adjacent nodes (Figure c). Note that this extension of entanglement is heralded, i.e. the knowledge that an entangling operation has been successful announces which modes in the quantum memories are entangled, thereby overcoming the probabilistic nature of photon transmission. The atomic excitation in those modes are then reconverted into photons, which are then subjected to another entangling operation that swaps the entanglement to the outside nodes (not shown). Obviously, in order to benefit from heralding, one photon per pair has to be stored in the quantum memory for the time it takes the other photon to travel half way towards the adjacent node, and the information about the successful entangling operation to travel back.}
  \label{figure_1}
\end{figure}

A possibility to overcome the problematic exponential scaling of loss and purity with distance is the quantum repeater, which was proposed by Briegel and coworkers in 1998 \cite{Briegel1998} and then further modified in subsequent years \cite{Duan2001,Childress2006,Simon2007,Jiang2007,Chen2007,Sangouard2007,Zhao2007,Choi2008,Sangouard2008}. The basic idea of a quantum repeater is to divide the long quantum channel into shorter segments and to distribute entanglement between end nodes of these segments. Then, the noisy entanglement is purified for each segment \cite{Bennett1996,Deutsch1996}, leading to one nearly pure pair per segment, and extended over adjacent segments by means of entanglement swapping \cite{Zukowski1993}. The purification procedure is repeated for the extended segments, and the whole protocol reiterated until high-purity entanglement is established between the end points of the link.

Quantum memories are essential in the repeater protocol as the initial distribution of entanglement as well as all purification steps are of probabilistic nature. Quantum memories allow to store entanglement, or purified entanglement in one segment until pure entanglement has also been established in the adjacent sections. Without quantum memory, all probabilistic steps would have to succeed at the same time.

Figures of merit for assessing the performance of a quantum memory for a quantum repeater include storage time, storage efficiency, and fidelity of storage (i.e. the similarity of the recalled quantum state with the input state). The storage time affects the maximum transmission distance, the efficiency determines the rate with which entangled states can be generated over long distances, and a high fidelity is essential for generating highly entangled states. Deriving minimum requirements for these parameters is difficult, as the best quantum repeater protocol (in terms of robustness against errors, and scaling of resources and communication time with distance) is still an active research topic (see e.g. discussion in \cite{Sangouard2008}). In order to find some benchmarks, let us consider a simplified, yet useful protocol that is inspired by \cite{Simon2007,Chen2007}. Note that it does not include purification, i.e. the maximum distance for quantum cryptography would be limited, similar to the quantum relay discussed in \cite{Collins2005}. Yet, it provides better performance than the quantum relay, which does not employ quantum memory.

As before, we divide the quantum channel into several short sections of length $L_0$ connecting adjacent nodes  (see Fig. 1). This time, each node contains 2$N$ sources of entangled photon pairs and two multi-mode memories, where each memory can store the quantum states encoded into $N$ photons. (The only exceptions are the first and last node at Alice and Bob, where only $N$ photon pair sources and one multi-mode memory is required.) One photon from each pair is sent into a quantum memory and stored, and the remaining photons are directed towards the two neighboring nodes, one on each side. Hence, $N$ photons are distributed in parallel into each direction, and $N$ times two photons meet half way between nodes, where they are subjected pairwise to an entanglement connecting operation (see section \ref{Tools}). This operation results in the establishment of heralded entanglement between specific memories modes in two adjacent nodes. The number of parallel channels $N$ is chosen such that the probability to entangle at least one pair of quantum memories modes per segment and round is close to one. Reconverting now the respective atomic excitations from those memory modes back into photons and making Bell state measurements finally allows entangling the memories at Alice's and Bob's. To keep the argumentation simple, we assume that the last Bell state measurement always succeeds, and that the photonic quantum states are not modified during storage in the memory (i.e. we assume the fidelity to be one). Assuming a segment length $L_0$ of at most 150 km for our simplified quantum repeater, the minimum requirement for the memory storage time $\tau_{min}$ thus is
\beq
\label{minimum storage time}
\tau_{min}=\frac{1}{c}L_0
\approx 1 ms.
\eeq
\noindent
where $c$ determines the speed of light in the communication medium.

Entanglement distribution over arbitrarily long distances obviously requires entanglement purification and thus relies on a fully implemented quantum repeater. The time required for establishing one pair of entanglement over 1000 km distance has been evaluated to be between ten seconds and thousands of seconds, depending on the implementation (see discussion in \cite{Sangouard2008}). Of particular interest for small communication times is the use of time-multiplexed multi-mode memories based on CRIB in rare-earth-ion doped crystals, as proposed in \cite{Simon2007}.

Another crucial property in addition to storage time is efficiency. Using again the simplified quantum repeater scheme, and assuming lossless quantum storage and recall, two photons at the end points of each segment become entangled after the first successful entanglement connecting procedure. However, for memory with limited efficiency $\epsilon$ (0$\leq\epsilon\leq 1$), this probability decreases to $P=\epsilon^2$. Without memory, where the distribution of entanglement would start from photon pair sources at the center between two nodes, the probability would be limited by transmission loss through the quantum channel: $P'=10^{-\textsl{a} L_0/10}$. Hence, we find quantum memory to be useful if the recall efficiency $\epsilon$ is larger than the transmission from one node to the center between nodes:
\beq
\label{minimum efficiency}
\epsilon_{min} > 10^{-\textsl{a} L_0/20}
\eeq
\noindent
For $L_0$=40 km and 150 km, we find $\epsilon_{min}$=0.4 and 0.03, respectively. To achieve reasonably high rates of entanglement generation over long distances, however, the detailed calculations in \cite{Childress2006,Simon2007,Choi2008} generally assume memory efficiencies larger than 90 \%.

Before we finish this section, we would like to point out that the absorption wavelength of the quantum memory is not particularly crucial for a quantum repeater. The only limitation stems from the requirement that sources of entangled photon pairs with one photon at the wavelength of the photon to be transmitted, and one at the atomic transition wavelength, must exist. This is likely to restrict the atomic transition used for absorption to wavelengths between a few hundred nanometers and a few micrometers. Note that the same (relaxed) limitation also holds for quantum repeater schemes '\`{a} la Briegel' \cite{Briegel1998}, where qubits encoded into photons are first transmitted, hence feature a wavelength that match the transparency windows in air or optical fibres, before being stored in quantum memories. Indeed, it is possible to set up a teleportation device preceding the quantum memory as to herald the arrival of the photon, and to teleport its quantum state onto a photon with different wavelength \cite{Marcikic2003}, the latter then being suitable for absorption. As before, the teleportation procedure requires a source of entangled photon pairs, whose properties impact on the possible atomic transition wavelengths. An alternative, non-heralded quantum information transfer between photons at different wavelengths that relies on non-linear up-conversion has been demonstrated in \cite{Tanzilli2005}.

\begin{figure}[h]
  \includegraphics[width=0.35\textwidth]{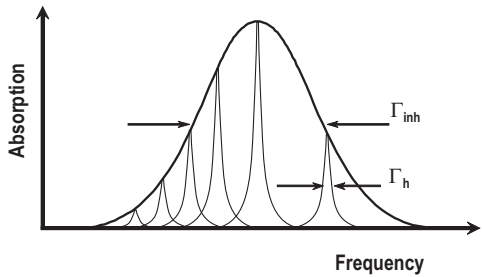}
  \caption{Illustration of the inhomogeneous linewidth  $\Gamma_{inh}$ for a resonant optical material and the homogeneous linewidth  $\Gamma_ h$ for individual groups of ions. Values of $\Gamma_ h$ as small as 50 Hz have been observed while $\Gamma_{inh}$ typically features values of 0.5 to 100 GHz.}
  \label{figure_2}
\end{figure}

\section{From data storage based on stimulated photon-echoes to quantum memory based on CRIB}
\label{History&CRIB}

\subsection{Historical development}
\label{History}

Quantum memories for time-bin based quantum communication should be able to store a photonic qubit (see Eq. \ref{qubit}) encoded into a photon in a superposition of being at two different positions, $x_0$ and $x_1$ at some given time $t$. Mathematically such a wave-packet can be expressed as

\bea\label
{time-bin qubit wavepacket}
\psi(x,t)=\nonumber\\
\bigg [\alpha S(\frac{x-x_0}{c}-t)+\beta e^{i\phi}S(\frac{x-x_1}{c}-t)\bigg ]e^{i(kx-\omega t)}
\eea
\noindent
where S(y) describes the shape of a basic wavepacket, $k$ is the wave vector, c is the speed of light and $\omega$ is the wave-packet angular frequency.

\bwt
\begin{center}
\begin{figure}[b]
  \includegraphics[width=.75\textwidth]{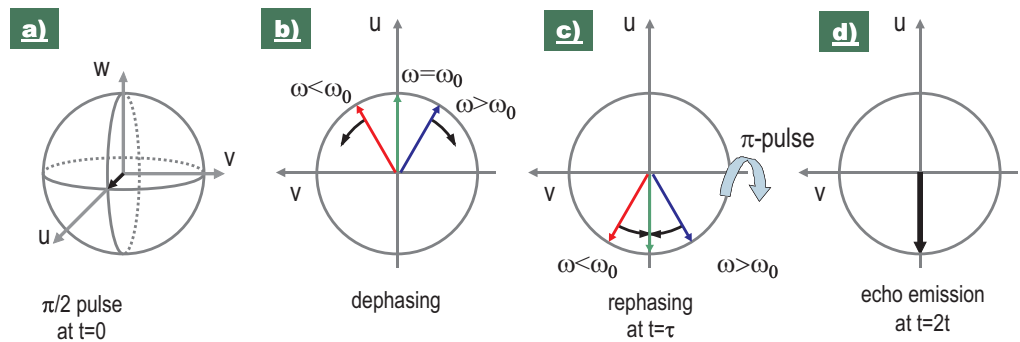}
  \caption{Illustration of the evolution of the Bloch vectors during the two-pulse photon-echo process. A first $\pi$/2 pulse rotates the Bloch vector from the negative w axis along the u direction (Figure a). The individual Bloch vectors precess freely in the uv plane and dephase, due to the inhomogeneous broadening of the transition (Figure b). At t=$\tau$, a $\pi$ pulse rotates all vectors around the v axis (Figure c). The Bloch vectors start rephasing, realign and build up a macroscopic coherence, and a photon-echo is emitted at time t=2$\tau$ (Figure d).}
  \label{figure_3}
\end{figure}
\end{center}
\ewt

Storage and recall of single photon states of the form given in Eq.\ref{time-bin qubit wavepacket} was discussed already 1993 by Kessel and Moiseev \cite{Kessel1993}. A specific experimental implementation was analyzed in 1998 \cite{Mohan1998} and demonstrated (for a sequence of equally prepared qubits) in 2003 \cite{Ohlsson2003}. Basically, the approach for storing and recalling single photon time-bin qubit states can be viewed as an outgrowth of the storage technique for classical optical data pulses put forward by Elyutin amd Mossberg in 1979 and 1982, respectively \cite{Elyutin1979,Mossberg1982}. Their idea, in its turn, was inspired by efforts for high density data storage that began in the middle of the 70's, (for a review see e.g. \cite{Moerner1988}). It was then proposed that the density of optical data storage could be increased beyond the diffraction limit by using a material where individual absorbers (atoms, ions, molecules, etc.) in the material absorb light with slightly different frequencies (see Fig. 2). In this way a light beam could be directed to one spatial point in the material and many bits could be stored at this location by simply changing the light frequency and in this way address different absorbers. Data bits were stored by promoting the absorbers to some excited state different from their ground state. In particular, it was realized that, in principle, any material with an inhomogeneously broadened absorption profile could be used for this purpose. By using the frequency dimension to address and store optical data, several thousands of data bits could be stored and addressed at a single spatial location, see e.g.\cite{Maniloff1995,Shen1997}. The maximum number of data bits that theoretically could be stored in a single location was then given by the ratio, $R$, between the inhomogeneous and homogeneous transition line-widths, $\Gamma_{inh}$ and $\Gamma_h$, respectively. Impressive material development lead to $R$ up to $10^8$ in Er$^{3+}$:LiNbO$_3$, and to an ultra-narrow homogeneous line width of only 50 Hz in Er$^{3+}$:\YSO (see \cite{Sun2002} and contribution by Y.C. Sun in \cite{RE_book}). In principle, it would then be possible to space different frequency channels for data storage by only about 100 Hz. However, plainly based on the uncertainty relation, addressing atoms at some frequency to read out a data bit stored at that frequency, without interacting with the ions in the next frequency channel 100 Hz away, would require optical pulses of several ms duration. The resulting sub-kHz data read and write rate does not seem very attractive.

\begin{figure}[h]%
\centering
\includegraphics[width=.45\textwidth]{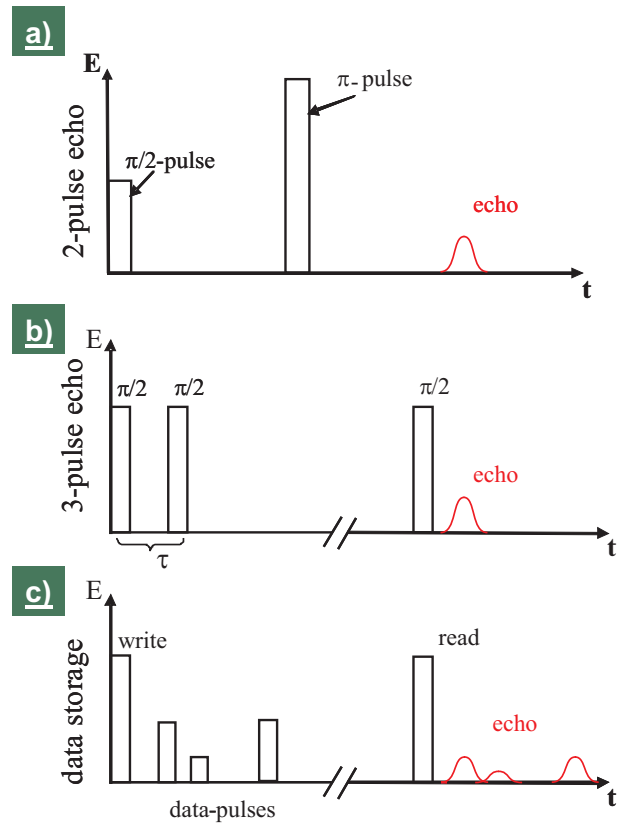}
  \caption{From two-pulse photon-echo to data storage. Figure a depicts the two-pulse echo, as explained in Fig. 3. Rephasing, i.e. echo emission can also be triggered by splitting the second ($\pi$) pulse into two $\pi$/2 pulses  (so-called three-pulse, or stimulated photon-echo, Fig. b). This can be understood when taking into account that the excitation pulses (pulse one and two) are represented in frequency space by a periodic function with period $2\pi/\tau$, where $\tau$ is the separation between the two pulses. This leads to a corresponding absorption grating in the medium, i.e. a periodic, frequency dependent variation of inversion. The spectrum of the third pulse is affected by this grating, resulting in two transmitted pulses instead of one, with the second one being the echo. Figure c shows the generalization of the three-pulse echo to data storage: The Fourier transformation of all input pulses (write pulse plus data pulses) is stored in the form of a spectral absorption grating. The spectrum of the read pulse is modified accordingly, giving rise to the emergence of a copy of the data pulses.}
\label{figure_4}
\end{figure}

Elyutin and Mossberg had, however, shown how to both eat the cake and have it using  photon-echoes \cite{Elyutin1979,Mossberg1982}. The photon-echo technique is a four-wave mixing technique, with pulses being separated in time. The three input pulses (two pulses in the case of the two-pulse echo) manipulate the system, putting the absorbers in the material in superposition states, which at some given time all oscillate in phase creating a macroscopic dipole moment (see Fig. 3 for a schematical description of a two-pulse echo). The output pulse (echo) is the radiation emitted from this temporally created macroscopic dipole moment. Specifically, to store a sequence of data, a brief preparation (write) pulse, $P_{write}$, with a pulse area equal to $\pi /2$ and a bandwidth equal or larger than the required data rate, is first sent into the sample (see Fig. 4). The data sequence, $P_{data}$, at a maximum data rate equal to the bandwidth of the pulse $P_{write}$, and duration shorter than the homogeneous relaxation time of the absorbers, is then sent into the material. The input radiation, determined by both write as well as data pulse, interacts coherently with the absorbers. As a result the frequency spectrum (Fourier transform) of the input data sequence (amplitude as well as phase) becomes imprinted as a frequency dependent modulation onto the sample absorption profile. Since both amplitude and phase are stored, the full temporal information of the light pulses is contained in the spectral interference patterns engraved in the sample, and later pulses can interact with these spectral gratings to perform temporal signal processing, including read-out. Mathematically, when the electromagnetic fields of the three input pulse sequences ($E_{write}(t)$, $E_{data}(t)$ and $E_{read}(t)$) are far from saturating the optical transition and the time separation between the pulse sequences are smaller than the transition relaxation time, the output field, $E_{echo}$, as a function of time, can be expressed as
\bea\label{classical echo field intensity}
E_{echo}(t)\propto
\int E_{write}^*(\omega )E_{data}(\omega)E_{read}(\omega)e^{i\omega t}d\omega
\eea
\noindent
where * stands for complex conjugate. In particular, if the frequency spectra of $E_{write}(\omega)$ and $E_{read}(\omega)$ are flat across the $E_{data}(\omega)$ bandwidth, $E_{echo}(t)$ will just be a copy of $E_{data}(t)$ (for a review see e.g. \cite{Mitsunaga1992a}). More complicated pulses $E_{write}(\omega)$ and $E_{read}(\omega )$ may be used for performing arbitrary operations on the data sequence, e.g., pattern or address recognition \cite{Harris1998}.

Consequently photon-echo techniques could be used to store and recall a pair of consecutive pulses, including amplitudes and the phase relation between the pulses. At least in retro-perspective it could then be a natural extension to assume that it would be possible to also store and recall a wave-packet superposition as that in Eq. \ref{time-bin qubit wavepacket}, as initially suggested by Kessel and Moiseev \cite{Kessel1993}.
However, the efficiency in a faithful storage and recall process of weak photon states using photon-echoes is in practice strongly limited. On the one hand, if the absorption is weak, most of the data pulse will just pass through the sample and the storage efficiency will be low. On the other hand, if the absorption is high the data sequence will be efficiently absorbed, but an efficient rephasing sequence using a $\pi$ pulse will invert the atomic medium and render it amplifying. It is then possible to generate echoes that are more intense than the data pulse \cite{Wang1999,Cornish2000,Crozatier2005}. However, the unavoidable amplified spontaneous emission generated in such a medium would add noise to the output echo field, making this approach inappropriate for quantum state storage. As mentioned before, the incompatibility of amplification and quantum state preservation is generally referred to as the no-cloning theorem \cite{Scarani2005}.

Yet, from a principle point of view, quantum mechanics states that for a closed quantum system $\ket{\psi (t_1)}$ at some time $t_1$, the system at some later time, $t_2$, will be described by $\ket{\psi(t_2)}=U\ket{\psi (t_1)}$, where $U$ is a unitary operator describing the evolution of the quantum state. Clearly the original state can be recreated by applying the inverse transformation $U^{-1}$ to $\ket{\psi(t_2)}$. This inspired ideas of the type where the recall process is just the time inversion of the storage process, which further may result in considering approaches like phase conjugation. For example, degenerate four-wave mixing where two counter-propagating pump beams effectively create a mirror that changes the sign of the wave vector of an incoming probe beam, causing the probe beam to return along its input path as if time propagated backwards \cite{Bloom1977}.

In the spirit of this idea, Moiseev and Kr\"{o}ll suggested in 2001 an approach to time-reverse the photon-echo storage process using atomic vapor as an inhomogeneously broadened lambda-type system \cite{Moiseev2001} (see Fig. 5 for a level structure). All atoms are prepared in an atomic state 1, e.g. some hyperfine ground state. Absorption of a single photon wave packet A, for instance of the form described by Eq.\ref{time-bin qubit wavepacket}, and with spectral width large as compared to the homogeneous line width, which is resonant with the inhomogeneously broadened $1\leftrightarrow 2$ transition, will create a nonzero probability amplitude for atoms to be in state 2. This probability amplitude would typically be distributed among around $10^9$ individual absorbers. Since the transition is inhomogeneously broadened, these dipole radiators will rapidly get out of phase, as is the case in the traditional photon-echo (see Fig. 3b). After some time $\tau$, shorter than the homogeneous dephasing time of the $1\leftrightarrow 2$ transition, a second pulse, B, collinear with the first pulse, resonant with the $3\leftrightarrow 2$ transition, and with a pulse area $\pi$ now transfers the probability amplitude in state 2 coherently to state 3. If the inhomogeneous broadening on the $1 \leftrightarrow 2$ transition is now reversed, the wave packet A can be recalled by sending in a third pulse, C, counter-propagating with the two first, resonant with the $2\leftrightarrow 3$ transition, and with pulse area $\pi$. Pulse C transfers the probability amplitude in state 3 coherently back to state 2. Due to reversal of the inhomogeneous broadening on transition $1\leftrightarrow 2$, all dipoles will start rephasing and a time-reversed replica of wave packet A will retrace its own input path in the backward direction at a time $\tau$ after the transfer of the probability amplitude from 3 to 2. As shown in the original paper, the efficiency of this storage and recall process in the absence of homogeneous dephasing processes can be 100\%. The requirement that the inhomogeneous broadening on the $1\leftrightarrow 2$ transition should be reversed may appear difficult to fulfil, but actually happens automatically for any Doppler-broadened transition, since the Doppler-shift in a gaseous sample has opposite signs for counter-propagating beams!

Gaseous media are, however, hampered of short storage times, determined by phase randomization of the emitted radiation due to atomic movement, velocity change, and loss of atoms from the interaction region \cite{Moiseev2004b}. A first adaption of the protocol towards solid state systems was suggested in 2003 \cite{Moiseev2003} using methods from nuclear magnetic resonance that allow quantum memory for microwave photons. In 2005 and 2006 three groups then described how the quantum memory scheme could be extended to solid state materials and photonic wave packets encoded into the optical part of the electromagnetic spectrum \cite{Nilsson2005a,Alexander2006,Kraus2006}. This protocol is now generally referred to as CRIB (see Fig. 6 for an illustration). It relies on spectral hole burning to create a narrow, isolated absorption line out of a broad, inhomogeneously broadened line, and employs reversible inhomogeneous broadening through externally controlled dc or ac Stark shifts, or Zeeman shifts.

\begin{figure}[t]%
\center
 \includegraphics[width=.25\textwidth]{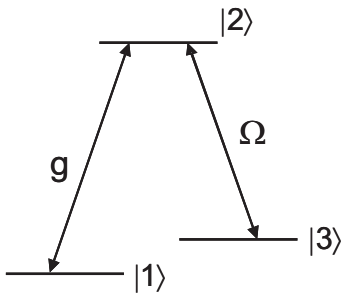}
  \caption{$\Lambda$-type atomic medium coupled to a quantum field with coupling constant $g$, and to a classical field with Rabi frequency $\Omega (t)$.}
\label{figure_5}
\end{figure}

As methods for creating an ensemble of absorbers that absorbed only at a specific frequency had already been developed \cite{Pryde2000,Nilsson2004,Crozatier2004,Rippe2005}, there was now an open path to develop long-lived quantum memories based on photon-echo techniques. Kraus and coworkers also discussed in a more general way what would be needed to perfectly store and recall quantum states (or classical optical data) in inhomogeneously broadened absorbers \cite{Kraus2006}. In particular, it was shown that direct spatial manipulation of the phase and frequency of the absorbers in the ensemble can cause a re-emission of the input pulse, A, without the need to use additional optical pulses, B and C. At this time the field had developed to a point where several groups had started work on actual experimental realizations as will be discussed in forthcoming sections.

\begin{figure}%
 \includegraphics[width=.37\textwidth]{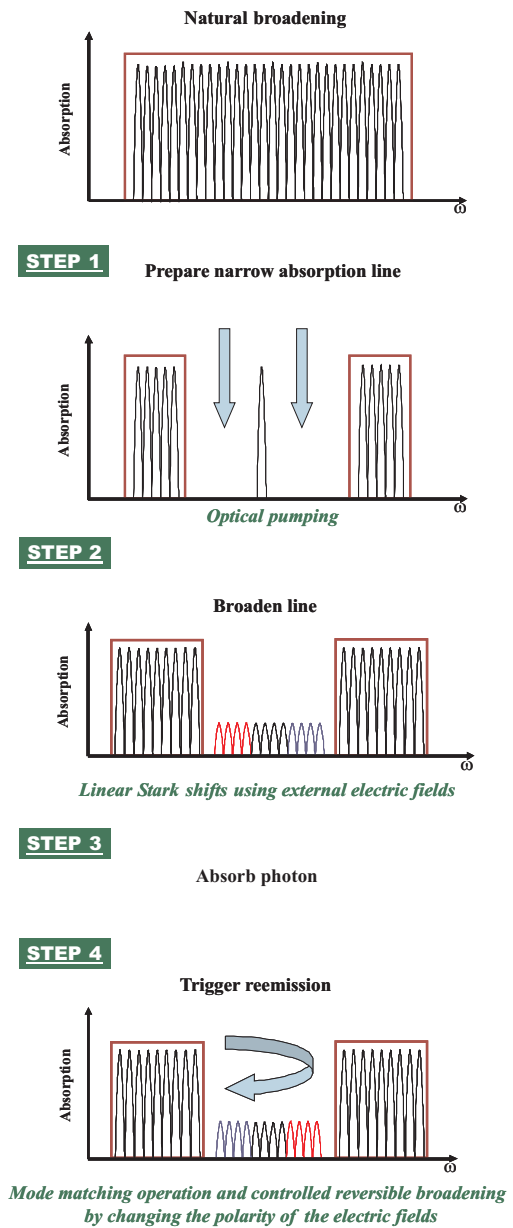}
  \caption{Quantum state storage based on CRIB in crystals with optical centers featuring permanent electric dipole moments. Starting with an ensemble of atoms or ions with broad, naturally (inhomogeneously) broadened absorption line, a narrow line is created through an optical pumping (or spectral hole burning) procedure (step 1). Next, the application of external electric fields results in a broadened absorption line with reduced optical depth (step 2). Note that this broadening is controlled and reversible, in contrast to the initial, natural inhomogeneous broadening. Then, a weak light field is directed into the medium and absorbed (step 3). Re-emission can be triggered at a later moment through the application of a $2kz$ position dependent phase shift for mode matching, and reversal of the inhomogeneous broadening (step 4). This leads to backwards emission of light in a time-reversed version of the input light field. The storage time is limited by the width of the isolated absorption line created in step 1, which is limited itself by the homogeneous linewidth $\Gamma_h$ (cf. Fig. 2).}
\label{figure_6}
\end{figure}

\subsection{Quantum memory based on CRIB}
\label{CRIB}

In the following, we present a simple theoretical description of the ideal CRIB protocol. It is based on a hidden symmetry in the equations that describe the absorption and retrieval of light, as discovered by Kraus \textit{et al.} in 2006 \cite{Kraus2006} (see also \cite{Sangouard2007}).

The resonant interaction of light with two level atoms has been treated in numerous textbooks (see e.g. \cite{Milonni&Eberly}). The evolution of the light electric field during propagation is determined by the atomic coherence $\sigma_{ge}(z,t,\Delta)$, according to Maxwell's wave equation
\beq
\Bigg (\frac{\partial^2}{\partial z^2}-\frac{1}{c^2}\frac{\partial^2}{\partial t^2}\Bigg )E(z,t)=\frac{n\wp}{\epsilon_0c^2}\int d\Delta G(\Delta)\frac{\partial^2}{\partial t^2}\sigma_{ge}(z,t,\Delta).
\label{wave_equation}
\eeq
\noindent
$G(\Delta)$ describes the inhomogeneous broadened line shape as a function of detuning $\Delta$ from the light carrier frequency $\omega_0$, and $n$, $\wp$, $c$ and $\epsilon_0$ are the atomic density, transition dipole moment, speed of light and permittivity in vacuum, respectively.

Assuming all atoms to be initially in the ground state, and the light propagating in the forward ($+z$) direction we can write the electric field and the excited atomic coherence as
\bea
\label{forward modes}
E(z,t)&=&E_f(z,t)e^{i(\omega_0 t-kz)}\\
\sigma_{ge}(z,t,\Delta )&=&\sigma_f(z,t,\Delta )e^{i(\omega_0t-kz)},\nonumber
\eea
\noindent
where we restrict ourselves to one transverse polarization mode for the electric field and atomic coherence. Assuming that $E_f(z,t)$ and $\sigma_f(z,t,\Delta )$ are envelope functions that vary slowly with $z$ and $t$, Eq. \ref{wave_equation} becomes

\beq
\Bigg ( c\frac{\partial }{\partial z}+\frac{\partial }{\partial t} \Bigg ) E_f(z,t)=i\beta\int d\Delta G(\Delta)\sigma_f(z,t,\Delta).
\label{Wave_equation_slowly_varying forwards}
\eeq
\noindent
where $\beta$ groups several, here irrelevant constants.
The associated change of the atomic coherence is described by the Bloch equations. Using again the slowly varying envelope functions (in rotating wave approximation), and assuming that most atoms in the medium remain in the ground state, these equations reduce to:
\beq
\frac{\partial}{\partial t} \sigma_f(z,t,\Delta)=-i\Delta\sigma_f(z,t,\Delta)+i(\wp/\hbar ) E_f(z,t).
\label{Bloch_equation_slowly_varying forwards}
\eeq
Note that the latter assumptions is certainly correct for a macroscopic number of atoms and weak light fields. However, reversible quantum evolution can also be shown for the more general case where the atomic inversion changes significantly  during absorption of the light field.

We assume that a weak pulse of light characterized by the (slowly varying) electric field $E_f(z,t)$ enters the optically thick atomic medium and is completely absorbed before time $t_0$. In other words, its field $E_f(t_0,z)$ has been completely mapped onto collective atomic coherence $\sigma_f$. For recall of the light pulse, we are interested in time reversal of the absorption process. Hence, we now look at the evolution of the backward propagating modes

\bea
\label{backwards modes}
E(z,t>t_0)&=&E_b(z,t)e^{i(\omega_0t+kz)},\\
\sigma_{ge}(z,t>t_0,\Delta )&=&\sigma_b(z,t,\Delta )e^{i(\omega_0 t+kz)}\nonumber,
\eea
\noindent which are initially empty, i.e. $E_b(z,t=t_0)=\sigma_b(z,t=t_0,\Delta)=0$. As a first step in the CRIB protocol, we have to transfer the excited atomic coherence $\sigma_f$ in a mode-matching operation to $\sigma_b$, which then becomes a source for an electric field propagating backwards.
This is done by introducing a physical, position dependent phase change of $2kz$, for instance by transferring the atomic coherences in a lambda system temporarily from the optical transition to a transition between closely spaced atomic ground states (see discussion at the end of section \ref{History}), or by exploiting a line shift that varies linearly along the propagation direction of light \cite{Kraus2006}. Next, we inverse the atomic detuning for all atoms $\Delta \rightarrow -\Delta$, which leads to the following light-atom equations:

\beq
\label{Wave_equation_slowly_varying backwards}
\Bigg ( -c\frac{\partial }{\partial z}+\frac{\partial }{\partial t} \Bigg )  E_b(z,t)=i\beta\int d\Delta G(\Delta)\sigma_b(z,t,\Delta)
\eeq
\noindent
and
\beq
\label{Bloch_equation_slowly_varying backwards}
\frac{\partial}{\partial t} \sigma_b(z,t,\Delta)=+i\Delta\sigma_b(z,t,\Delta)+i(\wp /\hbar) E_b(z,t).
\eeq

\noindent
Comparing this system of equations with the equations describing the absorption of the initial light pulse (Eqs.  \ref{Wave_equation_slowly_varying forwards} and \ref{Bloch_equation_slowly_varying forwards}), we find them to be identical provided we reverse the signs of the temporal derivatives ($\partial /\partial t\rightarrow - \partial /\partial t$) and the electric field in the latter two equations. This means that, after mode-matching and controlled reversal of inhomogeneous broadening, the atom-light equations describe a time reversed evolution of the envelope functions compared to the evolution during absorption. Hence, all atoms will finally be again in the ground state, and the light be re-emitted without loss into the backwards direction in a time-reversed version (we ignore the global phase change of $\pi$). In particular, for a time-bin qubit of the form given in Eq. \ref{qubit}, we thus find that the recalled state is associated with an exchange of the leading and trailing basic wavepackets.

Note that the electric field (as defined in Eq. \ref{backwards modes}) will accumulate a phase shift of $\phi=\omega_0\tau$ where $\tau$ is the storage time in the atomic medium. This phase shift becomes observable in the case of time-bin qubits, where the initially leading basic wave packet prevails longer in the medium compared to the initially trailing basic wave packet, see \cite{Moiseev2008b}. In conclusion, we find that the efficiency and fidelity (after correction of the deterministic changes of phase and order of basic wavepackets) of storage via CRIB in the here treated, ideal case must be one.

For the non-ideal case where re-emission is not a time reversed version of storage any more, it is obvious that symmetry arguments cannot suffice to assess these figures of merit, but that the evolution of the atom-light system has to be calculated in detail. Of particular interest are the cases where the photon is not always absorbed, and where light is recalled in forward direction, i.e. where the mode matching operation is not implemented. Interestingly, efficiency and fidelity depend on the type of inhomogeneous broadening in the atomic medium, where we distinguish between transverse, and longitudinal broadening. In transverse broadening, the atomic absorption line is equally inhomogeneously broadened for each position $z$, i.e. a controlled line shift is applied transverse to the propagation direction of light. Longitudinal broadening
refers to atomic media where the absorption line for each position $z$ is narrow, but varies monotonically through the medium: $\Delta=\chi z$. This requires controlled line shifts in longitudinal direction. As mentioned before, for sufficiently large optical depth, the efficiency is unity, regardless the type of broadening.

For the case of transverse broadening and limited optical depth $\alpha L$, where $\alpha$ is the absorption coefficient in cm$^{-1}$ (not to be confused with the coefficient $\textsl{a}$, given in db/cm, used to calculate the transmission through an optical fibre in section \ref{Quantum repeater}) and $L$ the length of the medium in cm, the recall efficiency $\epsilon_{b}$ in backwards direction has been derived independently in \cite{Moiseev2004,Sangouard2007}, and is given by:
\bea
\label{recall efficiency transverse backwards}
\epsilon_b^{(t)}=(1-\exp\{-\alpha L\})^2.
\eea
\noindent
As shown above, the maximum efficiency for backwards recall is one (assuming large optical depth). For recall in forwards direction, the efficiency is given by
\bea
\label{recall efficiency transverse forwards}
\epsilon_f^{(t)}=(\alpha L)^2\exp\{-\alpha L\},
\eea
\noindent
as demonstrated in \cite{Sangouard2007}. In this case, the efficiency is limited to 54\% for an optical depth of $\alpha L$=2.
Regardless the direction of recall and except for the exchange of the leading and trailing basis wave packets and deterministic phase change, the recalled quantum state exactly resembles the input state.

For longitudinal broadening where the resonance frequency varies monotonically as a function of position, the efficiencies are given by
\bea
\label{recall efficiency longitudinal}
\epsilon_b^{(l)}=\epsilon_f^{(l)}=(1-\exp\{-2\pi\kappa/\chi\})^2,
\eea
\noindent
where $\kappa$ groups several atomic parameters, and $2\pi\kappa /\chi$ characterizes the effective optical depth $(\alpha L)_{eff}$ of the medium, see \cite{Moiseev2008a,Longdell2008}. Note the equivalence with the efficiency of backwards recall in the transverse broadened case (Eq. \ref{recall efficiency transverse backwards}). Hence, we see that even though emission in forwards direction is at odds with perfect time reversal, the efficiency (for large effective optical depth) can still be one! However, the photonic wave packet recalled in forward direction is associated with a frequency chirp, i.e. the recalled quantum state does not fully resemble the input state, even after having compensated for the exchange of wavepacket ordering and storage time related phase \cite{Moiseev2008a}.

\section{Material considerations}
\label{Material considerations}
The development of quantum memory depends on the availability of an atomic medium with appropriate properties, regardless the specific protocol pursued. Fortunately, as briefly addressed in section \ref{Quantum repeater}, the absorption wavelength is not particularly crucial in the case of a quantum repeater, which makes many different atomic media potentially suitable.

Material requirements for CRIB-based quantum memory (see Fig. 6) include the possibility to inhomogeneously broaden an optical transition in a controlled way. The induced broadening needs to be large compared to the inherent broadening of the transition, and it should be possible to reverse it in a time which is short compared to the inverse of the inherent broadening. Extended requirements to allow building of a useful quantum repeater include a long storage time, which is limited by the homogeneous linewidth of the optical transition (the 1$\leftrightarrow$2 transition in Fig. 5) or the transition used to store the quantum state (the 1$\leftrightarrow$3 transition), and spectral diffusion within the ensemble. In addition, the spacing to neighboring ground or excited state levels (e.g. other hyperfine levels) should be large compared to the spectral width of the photons to be stored, so that the light interacts only with two-level systems comprised of one ground and one excited state. Finally, for efficient storage, atomic ensembles with large optical depth are required.

\subsection*{Rare-earth-ion doped solids}

Currently, research into CRIB-based quantum state storage focuses on rare-earth-ion doped solids (RE doped solids) as material candidates, and first proof-of-principle demonstrations have recently been reported \cite{Alexander2006,Alexander2007,Hetet2008}. RE solids have been studied for more than half a century, in particular due to their interest for solid state lasers (see e.g. \cite{Powell1998}) and fibre optics amplifiers \cite{Desurvire2002,Becker1999}, but more recently also for laser stabilization to programmable frequency standards \cite{Sellin1999,Strickland2000, Sellin2001,Cone2001,Pryde2002,Boettger2003,Boettger2007}, and radio frequency analyzers \cite{Colice2006,Gorju2007,Mohan2007}.  These studies, together with extensive fundamental investigations (see \cite{Macfarlane1987a,Sun2002} and contributions by G. Liu and Y.C. Sun in \cite{RE_book}), have resulted in broad understanding of interactions in RE solids. Even though the properties required for quantum state storage differ from those required for other applications, one can hope that this knowledge, paired with an almost uncountable number of possible RE solids, will eventually lead to novel materials that remove shortcomings in those currently investigated.

Rare earth metals doped or implanted into inorganic solid state hosts (both crystals and glasses) generally form trivalent (triply charged) ions.
They are set apart from other transition-metal ion materials, since their optically active 4f$^N$ electrons are tightly-bound and shielded by outer 5s$^2$ 5p$^6$ closed shells, giving rise to atomic-like character for the 4f$^N$ levels even in a crystalline solid at doping densities as great as 10$^{18}$/cm$^3$. Transition wavelengths range from the infrared to the ultraviolet. Currently, many investigations focus on Thulium, Erbium, Europium, Neodymium, or Praseodymium doped crystals, often \YSO or LiNbO$_3$, and on transition wavelengths (which are largely determined by the specific rare-earth-ion transition, not the host) around 795 nm, 1532 nm, 580 nm, 880 nm and 606 nm, respectively.

Work over several years by the Cone group has shown that sample-to-sample variations in crystal properties can be significant, affecting both static and dynamic properties. One must be cautious in drawing conclusions based on a single sample or single doping concentration.

\subsection{Energy levels}
The 'free ion' levels of a rare earth ion are modified by a weak interaction with the host crystal lattice. Electrostatic interactions, covalency, and overlap with neighboring ligands and other host ions are typically well understood and are usually described by single particle operators called the 'crystal field', see \cite{Dieke1963,Judd1963,Wybourne1965,Huefner1978,Newman2000} and contribution by G. Liu in \cite{RE_book}. The degeneracy of J-multiplets is raised, consistent with the local RE site symmetry, giving rise to 'crystal field manifolds' of levels typically spread over a few hundred cm$^{-1}$.
Depending on the number of remaining 4f electrons, even or odd, RE ions form so-called non-Kramers, or Kramers ions, respectively. For non-Kramer ions, the J-degeneracy can be and usually is completely lifted, while a two-fold degeneracy or greater remains in the case of Kramer ions.
Additional structure arises from magnetic and electric hyperfine interactions $\textbf{H}_{hfs} = A_J \textbf{I}\cdot \textbf{J} + P [(I_Z^2 - I(I+1)/3) + \eta(I_X^2 - I_Y^2)/3]$, where $A_J$ is a familiar atomic constant and $P$ and $\eta$  describe the nuclear electric quadrupole interaction (when pseudo-spin representations are used, as is typically the case, these interactions and Zeeman interactions can look different as a result of anisotropy imposed by the nature of the electronic states). Level spacing between hyperfine levels ranges from a few tens of MHz in the case of Europium and Praseodymium  to the GHz scale for Terbium and Holmium doped crystals \cite{Dieke1964,Wybourne1965,Huefner1978,Macfarlane1987a}. In some systems the $f$ electrons also interact with nuclei of surrounding ligands; this superhyperfine or transferred hyperfine interaction \cite{Macfarlane1987a} is a factor in both level structure and spin dynamics. The hyperfine and superhyperfine level structure can differ quite dramatically from one system to another, and these splittings and structure may impose a limit on the spectral bandwidth of the photons to be stored.

\subsection{Homogeneous linewidth}

The shielding of the 4f electrons reduces electron orbit-lattice interaction as compared to other solids, thereby minimizing the effect of dynamic perturbations by phonons. This results in transition intensity concentrated in narrow zero-phonon lines, often with no obvious intensity appearing in phonon sidebands.
The sharpest optical transitions between the $4f^N$ states arise from the ground level of the lowest manifold to the lowest level in an excited manifold. Transitions to higher levels in each manifold are typically broadened by non-radiative cascade decay to lower levels within the manifold, or in some cases to lower manifolds nominally lying within a few phonon energies. The homogeneous linewidth $\Gamma_h$ of RE solids is temperature dependent, with thermal broadening arising from coupling to phonons, spins, and, in the case of glasses, a broad distribution of low-frequency tunneling modes (two-level systems or TLS) \cite{Huber1984,Broer1986,Geva1997}. Below $\approx$ 4 K contributions from phonon absorption and emission, and phonon Raman scattering are usually negligible, except in cases where the crystal field splitting to the first level of the manifold is small, on the order of a few times $kT$. For crystalline hosts, this results in linewidths of typically around a few kHz, but values as small as 50 Hz have been observed in Eu$^{3+}$:\YSO
and Er$^{3+}$:\YSO, corresponding to less than one part in $10^{12}$ of the transition frequency \cite{Equall1994,Boettger2003}. Linewidths in glasses or optical fibres are generally larger, due to coupling between the RE-ion and TLS; yet, widths below 1 MHz have been observed in Er$^{3+}$ doped fibers and glasses at very low temperature and high magnetic fields \cite{Staudt2006,Macfarlane2006,Sun2006}.

In addition to coupling to phonons and two-level systems, there is a variety of other mechanisms that can increase the homogeneous linewidth.
Radiative and non-radiative decay make a familiar contribution usually discussed in terms of a lifetime $T_1$ \cite{Ermeneux2000}, but metastable levels, can have T$_1$ values up to 1 - 10 ms, giving lifetime-limited values as small as $\Gamma_h$=10-150 Hz. For excited levels within a manifold, on the other hand, non-radiative decay times can be in the nanosecond to picosecond range. Nuclear and electronic spin fluctuations make especially material-dependent contributions to  $\Gamma_h$. Unlike ions with even numbers of electrons, those with odd numbers are required by Kramer's Theorem \cite{Wybourne1965} to have degenerate electronic levels in the absence of a magnetic field. All those ions are paramagnetic and thus sensitive to fluctuating local fields and to applied magnetic fields. Magnetic contributions to dephasing can be avoided in part by choice of crystal composition, ion dilution, or applied magnetic field. Indeed, induced Zeeman splittings both small and large can dramatically affect spin dynamics and reduce spin contributions to decoherence; the resonance required for nuclear spin flip-flops can be disrupted by small magnetic fields, and in paramagnetic materials containing Er$^{3+}$ or Nd$^{3^+}$ ions, large applied magnetic fields can be used to freeze out electronic spin fluctuations
\cite{Boettger2003,Macfarlane2004,Boettger2006a,Sun2002,Hastings2008}.
Furthermore, when RE ions are optically excited their interactions with neighbors change, introducing frequency shifts resulting in instantaneous spectral diffusion; the term excitation-induced frequency shifts is also applied \cite{Liu1987,Liu1988,Huang1989,Liu1990,Kroell1991,Mitsunaga1992b,Equall1994,Altner1995,Altner1996a,Altner1996b,Graf1998}. This line broadening mechanism is not familiar from other areas of optical physics.

An additional important dynamical process is traditional spectral diffusion, which results from time-dependent perturbations of each ion's transition frequency due to the dynamic nature of the ion's environment. The accumulating frequency shifts cause each ion to undergo a limited random walk in frequency, or to diffuse, through the optical spectrum, leading to an apparent line broadening with time, hence a progressive increase in the rate of phase decoherence. For RE materials at low temperatures, an important mechanism for spectral diffusion is the magnetic dipole-dipole interaction of each optically active ion with the other electronic and nuclear spins in the host material. As discussed above, this can be particularly important for Kramers ions such as Erbium \cite{Boettger2003,Boettger2006a}.
Macfarlane and Shelby \cite{Macfarlane1987a} reviewed many results showing that for even-electron systems, nuclear spin dynamics in the host material, especially flip-flop transitions, are the dominant mechanism for spectral diffusion; for even-electron systems magnetism is far weaker but not absent.

The development and characterization of materials with narrow optical lines suitable for quantum information device concepts has been a continuation of the extensive studies reviewed by Macfarlane and Shelby \cite{Macfarlane1987a}, Macfarlane \cite{Macfarlane2002,Macfarlane2007}, and Sun \cite{Sun2002,RE_book}. It was realized that to reduce decoherence one should reduce interactions of an ion with its crystal surroundings. Extremely long optical coherence times, up to 4 ms, have been achieved by choosing constituent elements of the host material to have small nuclear magnetic moments or small natural abundance of magnetic isotopes. Initial attention focused heavily on the non-Kramer ions Eu$^{3+}$, Pr$^{3+}$, and Tm$^{3+}$ where an even number of electrons can give singlet electronic crystal field levels that to first order have no electronic magnetic moment, though nuclear magnetic moments are still present. Using this strategy, kilohertz homogeneous linewidths  were reported for Eu$^{3+}$:\YSO \cite{Macfarlane1981, Babbitt1989}, and later linewidths approaching 100 Hz were measured in Eu$^{3+}$:\YSO \cite{Yano1991,Equall1994,Koenz2003,Macfarlane2004}. Kilohertz widths were also achieved in Pr$^{3+}$:\YSO \cite{Equall1995} and Er$^{3+}$:\YSO \cite{Boettger2003b,Macfarlane2004,Boettger2006a,Boettger2006b,Boettger2008}, where the narrowest currently reported homogeneous linewidth in any material of 50 Hz was observed.

From these studies, one can conclude that millisecond storage times may be achievable in RE crystals, which would already be interesting for the simplified quantum repeater discussed in section \ref{Quantum repeater}. However, second-long storage times, as probably required for the full quantum repeater and quantum communication over distances exceeding 1000 kilometers, seem currently difficult to realize for optical transitions. Longer coherence times can be expected for hyperfine ground state superpositions when applying magnetic fields suitable to remove the sensitivity of the transition to magnetic fields to first order \cite{Longdell2006}. Using this approach, coherence times of 82 ms have been reported for Pr$^{3+}$:\YSO \cite{Fraval2004}. This value has been further improved to more than 30 sec using dynamic decoherence control \cite{Fraval2005}.

Three-level lambda systems may not always be accessible, however, depending on selection rules, even if an otherwise suitable level structure exist.
The possibility to improve branching ratios towards a second ground state has been investigated theoretically for Pr$^{3+}$:LiYF$_4$ \cite{Goldner2008} and Tm$^{3+}$:Y$_2$O$_3$ \cite{Chaneliere2008}. Furthermore, extensive calculations and measurements were carried out to demonstrate that Nd$^{3+}$:YVO$_4$ \cite{Staudt2008} and Tm$^{3+}$:YAG \cite{Guillot-Noel2005,de-Seze2006} can form suitable lambda systems, and measurements of branching ratios and nuclear spin coherence lifetimes have been reported recently for the latter RE crystal \cite{Louchet2007,Louchet2008}.

It is interesting to note that the use of lambda systems not only promises increasing the storage time. In addition, transferring the atomic coherence between different levels using counter-propagating $\pi$-pulses also allows realizing the $2kz$ mode-matching operation required for CRIB, see e.g. \cite{Moiseev2004,Nilsson2005a,Kraus2006}.

\subsection{Inhomogeneous broadening}

Currently the major obstacle for RE solids to be used for CRIB is the inhomogeneous broadening of the narrow optical transitions. Typically inhomogeneously broadening $\Gamma_{inh}\approx$ 0.5-100 GHz arises due to local strains and defects in the crystal structure, see Fig. 2. This broadening is similar to Doppler broadening in gases. For infrequent cases, inhomogeneous widths in crystals of $\Gamma_{inh} \approx$ 10 MHz are possible \cite{Macfarlane1992, Macfarlane1998,Thiel2008c}, which would, however, still limit the storage time to few tens of nanoseconds if one were to perform CRIB without preparation steps. Efforts are under way to achieve much narrower distributions \cite{Thiel2008c}.

In order to form narrow, isolated absorption lines (spikes), the inhomogeneously broadened line must be manipulated through so-called spectral hole burning techniques (SHB) \cite{RE_book}. This allows optical pumping or coherent transfer of individual subgroups of ions with resonance frequencies near the desired spike to other atomic levels that do not participate in the procedure for quantum state storage  \cite{Pryde2000,Nilsson2004,Crozatier2004,Rippe2005,Wesenberg2007,Alexander2006,Hetet2008,Hastings2008,Rippe2008}. This technique has allowed creating isolated lines of a few tens of kHz, which then limit the storage time to a few hundred $\mu$sec \cite{Sangouard2007}.

\subsection{Stark shifts}

The ion energy levels can be manipulated using the Zeeman and Stark effects through the Hamiltonian
$\textbf{H}= -[\mu_B(\textbf{L}+g_s \textbf{S})]\cdot \textbf{B}-\bf{\mu_n B}-\textbf{p}\cdot \textbf{E}$,
where  $\mu_B (\textbf{L} + g_s \textbf{S})$ and $\textbf{p}$ are the static magnetic and electric dipole moments of the $4f^N$ electrons, $\textbf{B}$ and $\textbf{E}$ are applied magnetic and electric fields, and  $\bf{\mu_n}$ is the nuclear magnetic moment of the optically active ion. Of particular interest for CRIB is the linear Stark effect, which can be observed for RE impurities in low symmetry sites. Provided the permanent electric dipole moments of the states coupled by the optical field are different, it leads to a shift in resonance frequency (see a recent review by Macfarlane \cite{Macfarlane2007}), which can be exploited for controlled reversible inhomogeneous broadening of the associated transition.

The magnitude of the Stark shift depends on the particular transition, the crystal host, and the orientation of the electric field with respect to the permanent dipole moment difference. The largest Stark shifts are of the order of 100 kHz/Vcm$^{-1}$\cite{Macfarlane2007}. In RE doped crystals, where the application of a dc electric field leads to a shift or discrete splitting of an absorption line, controlled inhomogeneous broadening of the order of several hundred MHz can potentially be induced by applying an electric field gradient with field strength varying between -1000 Vcm$^{-1}$ and +1000 Vcm$^{-1}$. In RE doped glasses, where the orientation and magnitude of dipole moments varies randomly, a constant electric field suffices for broadening.

Of particular interest in the context of Stark broadening are waveguiding structures, which, due to the possibility to space electrodes closely, allow broadening of hundreds of MHz with voltages of only a few tens of volts \cite{Hastings-Simon2006}. Furthermore, the use of LiNbO$_3$ waveguides (which are used in the telecommunication industry for phase and intensity modulators operating at more than 10 Gbps) allows tailoring electrodes on demand, and switching electric fields of several hundred Vcm$^{-1}$ in a fraction of a nanosecond.

\subsection{Absorption}

The $4f^N \leftrightarrow 4f^N$ optical transitions are 'forced' electric dipole transitions that arise due to small admixtures of excited configurations into the $4f^N$ states by odd parity terms in the crystal field, see \cite{vanVleck1937,Judd1962,Ofelt1962} and contribution by M.F. Reid in \cite{RE_book}. Selection rules are  $\Delta S = 0$,  $\Delta L\leq 6$, and $\Delta J\leq 6$, though intermediate coupling in most excited states usually means that S and L are not especially good quantum numbers. In cases where the selection rule  $\Delta J\leq 1$ is satisfied, allowed magnetic dipole transitions are also observed. Those transitions are in the infrared and have become accessible with diode lasers over the past ten years. Magnetic dipole transitions are particularly important for Er$^{3+}$ materials including Er$^{3+}$:\YSO or Er$^{3+}$:LiNbO$_3$ \cite{Boettger2003,RE_book,Thiel2008}. General symmetry considerations for optimizing the Rabi oscillation in solids of any symmetry have been published by Sun \textit{et al.} \cite{Sun2000}.

The forced electric dipole transitions between the nominally $4f^N$ states have oscillator strengths $f$ with typical values ranging from $10^{-6}$ to $10^{-8}$. Larger oscillator strengths at the upper end of this range have been found for Nd$^{3+}$:YVO$_4$($f$=8 x 10$^{-6}$) \cite{Sun2002}, Er$^{3+}$:LiNbO$_3$ ($f$=1 x 10$^{-6}$) \cite{Sun2002,RE_book,Sun2008b,Thiel2008a}, and Tm$^{3+}$:LiNbO$_3$ ($f$=3 x 10$^{-6}$) \cite{RE_book,Sun2008b,Mohan2007,Thiel2008a}. While these numbers are still small, the ultra-narrow linewidths and high number densities readily allow optical depths of 1-10 in mm scale samples \cite{Hastings2008,Hedges2008}. Note that the optical depth before controlled broadening has to be very large to ensure sufficient absorption after broadening. It can be further increased by using long waveguiding structures, or multi-pass configuration.

\begin{figure}[t]
\centering
\includegraphics[width=.45\textwidth]{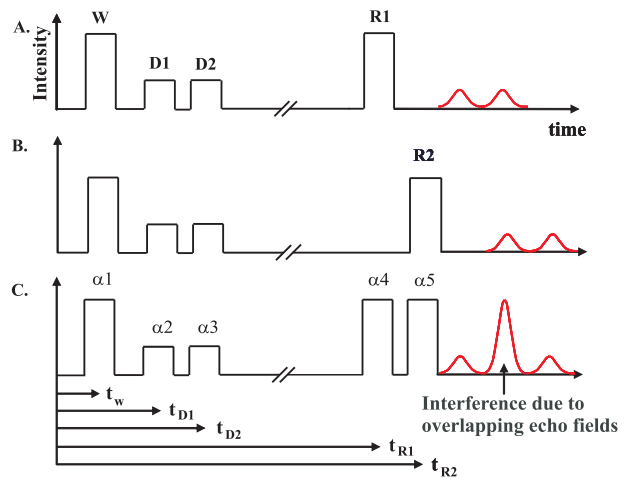}
 \caption{Pulse sequence for interfering time-bin pulses stored in a solid-state memory based on a Erbium-doped LiNbO$_3$ crystal. A strong write pulse W prepares the medium for storing the time-bin pulses D1 and D2. Two strong read pulses R1 and R2 result in photon-echo emissions which lead to an interference in the central time-bin. Interference fringes could by produced by varying one of the phases $\alpha2$ to $\alpha5$ (cf. Fig. 8).} \label{figure_7}
\end{figure}

\section{Stimulated photon-echo as a test-bed for quantum memory}
\label{photon echoes as test bed}

The implementation of CRIB is currently still challenging. Interestingly, it is possible to examine many features of CRIB using much simpler traditional photon-echoes \cite{Staudt2007a,Staudt2007b}, as both approaches to storage are based on re- and dephasing of coherences in an inhomogeneously broadened medium.

The performance of a quantum memory can be qualified by measures such as memory time, efficiency, and fidelity (see discussion in section \ref{Quantum repeater}).
The efficiency is generally defined as the total
probability of absorbing a photon and later re-emitting it (on
demand) in the chosen temporal and spatial mode. We define the memory fidelity $F$ as the overlap
between the state of the photon before storage (input) $\ket{\psi_{in}}$ and the
state of the photon after storage (output), which is generally mixed and denoted by a density matrix $\rho_{out}$: $
F=\bra{\psi_{in}}\rho_{out}\ket{\psi_{in}}$.
In the case of quantum
information applications using single-photon states such as qubits (Eq. \ref{qubit}), one can
post-select the cases when a photon was actually re-emitted from
the memory, in which case the normalized (or post-selected)
fidelity is independent of the efficiency of the memory. In this
section we will consider post-selected fidelities. The fidelity of
the memory will then be lowered by processes that destroy the
phase coherence of the two qubit basis state $\ket{0}$ and $\ket{1}$, or modify the probability
amplitudes $\alpha$ and $\beta$ in a non-deterministic (i.e. non-reversible) way.

Storage and recall of multi-photon data pulses via stimulated photon-echoes has been studied for decades (see section \ref{History}), with emphasis on unperturbed recall of time varying optical power (which defines the data pulses). However, coherence properties have primarily received attention only in connection with erasure of data \cite{Akhmediev1990,Arend1993,Elman1996,Dyke1999}, as phase coherent storage is of no concern in classical communication. Investigation in view of the requirements of quantum communication have started only very recently. The experiments we will review here, reported in \cite{Staudt2007a,Staudt2007b}, were based on photon-echoes in the classical pulse regime. Hence bright coherent states
of light were stored and retrieved from RE solids using either ordinary (two-pulse) or stimulated (three-pulse) photon-echoes. As we will show, this allows drawing important conclusions about the phase coherence of quantum memories based on RE solids.

\begin{figure}[b]
 \centering
  \includegraphics[width=.45\textwidth]{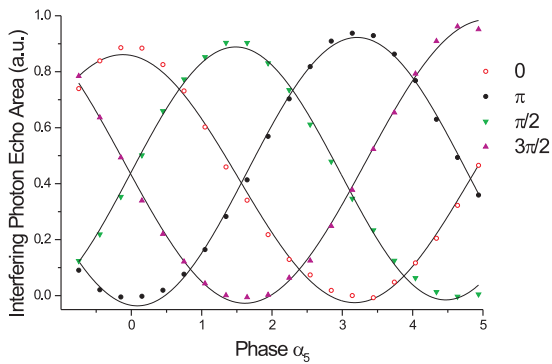}
 \caption{Four classical states ($\alpha2,\alpha3$)=(0,0), (0,$\pi$/2), (0,$\pi$), (0,3$\pi$/2), analogous to the quantum states used in the BB84 quantum cryptography protocol \cite{Bennett1984,Gisin2002}, are stored, retrieved and analyzed (by scanning the phase $\alpha5$) with close to 100$\%$ fidelity. This is possible even though the probability of retrieval from the memory is only a few percent, limited by the efficiency of the photon-echo process and by decoherence processes in the storage material.}
 \label{figure_8}
\end{figure}

\subsection{Interference of two subsequent wave-packets recalled from one crystal}

The experiment reported in \cite{Staudt2007a} demonstrated that information encoded in the amplitudes and phases of two subsequent coherent pulses can be stored in and retrieved from a single RE solid with close to 100$\%$ fidelity. This way of encoding information is common in quantum key distribution systems where time-bin qubits are used (see e.g. \cite{Tittel2001}). In a time-bin qubit the photon is in a coherent superposition of being in an early and late time bin. This superposition can be created by sending a single photon through an un-balanced Mach-Zehnder (MZ) interferometer. The time-bin qubit has the advantage of being robust with respect to depolarization effects in optical fibers and can be sent over long distances without any need for active polarization control of the fiber link. The analysis of the time-bin qubit is normally done by projection measurements using another MZ interferometer. In this experiment, the two coherent pulses, or time-bin pulses, can be considered a classical equivalence to a time-bin qubit.

The time-bin pulses were created by tailoring a continuous wave laser at 1.53 $\mu$m wavelength with large coherence length using a combined telecommunication intensity and phase modulator. The pulses were then stored and retrieved from a solid-state memory consisting of Erbium ions doped into a LiNbO$_3$ crystal with a single-mode channel waveguide on its surface, where the Erbium ions have a transition at 1.5 $\mu$m in the telecommunications window (more details about the crystals can be found in Ref. \cite{Staudt2007a}). The memory process was based on stimulated photon-echoes (SPE), see Figs. 3 and 4. In the most simple optical storage experiments using SPE the pulse sequence consists of a strong write pulse, a weak data pulse and a strong read pulse, see Fig. 7a. The write/read pulses are ideally $\pi$/2 pulses while the data pulse should be sufficiently weak ($\ll$ $\pi$/2) such that the echo is a linear transformation of the data pulse. If the data pulse consists of the two time-bin pulses as discussed above, the SPE process will generate an echo consisting also of two coherent pulses, as shown in Fig. 7a. By sending in two strong read pulses, one can trigger two partial retrievals of the stored time-bin pulses, separated by the time difference between the read pulses. If this time difference is equivalent to the separation between the two time-bin pulses, the second time-bin pulse retrieved by the first read pulse will overlap with the first time-bin pulse retrieved by the second read pulse and interfere, see Fig. 7c. The interference depends on the phases of the stored time-bin pulses
($\alpha2$ and $\alpha3$) and the two read pulses ($\alpha4$ and $\alpha5$). A complete interference fringe can be obtained by scanning one of the phases, which is shown in Fig. 8. For the four input states with different phases $\alpha3$, the observed  visibility $V$ was always close to 100$\%$, corresponding to a fidelity $F=(1+V)/2$ close to 100$\%$. These results demonstrate that the relative phase and amplitude ratio of time-bin pulses can be preserved during storage in the optical memory, with close to perfect fidelity\footnote{In
principle, one could create the time-bin pulses using a MZ
interferometer, store them in a RE solid using SPE, and then
analyze the retrieved time-bin pulses using another MZ
interferometer. The main difficulty using that approach would have been the stabilization of the MZ interferometers with about 12 m path length difference, as the time-bin pulses were separated by 60 ns.}.

\subsection{Interference of wave-packets recalled from two crystals}

The second experiment \cite{Staudt2007b} that we will review here aimed at demonstrating phase coherent storage of optical pulses in two independent and spatially separated solid state memories. This property is important when storing the two basis qubit states $\ket{0}$ and $\ket{1}$ in different memories, and is also a requirement for quantum repeater protocols that are based on interference of quantum states of light retrieved from different quantum memories (see discussion in \cite{Sangouard2008}). The experiment reported in Ref. \cite{Staudt2007b} was designed with
these two aspects of quantum repeaters in mind.

The experimental set-up is shown in Fig. 9a. The two solid-state memories were two Erbium-doped LiNbO$_3$ crystals, both with single-mode channel waveguides, that were placed in a fiber-based equilibrated MZ interferometer. A bright optical pulse is sent into the interferometer and is partly absorbed in the memories. The absorbed photons are then stored as a coherent superposition of atomic excitations in the two memories. After some time, a second bright pulse is sent into the interferometer. It serves as a read (i.e. rephasing) pulse in the photon-echo experiment. The retrieved pulse (photon-echoes) will then interfere at the output of the interferometer, on the condition that the storage in the two spatially and independent memories is phase coherent. To observe interference fringes at the output, the path length difference of the interferometer was tuned by applying a variable voltage to a piezo element that slightly pulled the fiber in one of the arms. As shown in Fig. 9b, high-visibility interference fringes were observed while scanning the phase. These results clearly demonstrate that optical pulses can be stored as a coherent superposition in two independent solid state memories using photon-echo type processes.

\bwt
\begin{center}
\begin{figure}[t]
 \includegraphics[width=.95\textwidth]{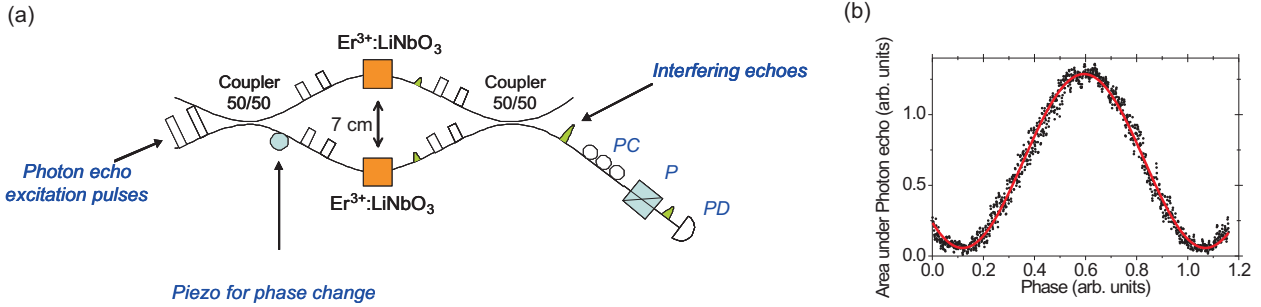}
 \caption{(a) Two Er$^{3+}$:LiNbO$_3$ waveguides cooled to below 4 K are placed in the arms of a fiber-optic interferometer. The excitation light pulses are sent through the interferometer and the generated echoes interfere at the second coupler. In order to project the polarizations onto one axis we used a polarization controller (PC), a polarizer (P) and a photo detector (PD). (b) An interference fringe as a function of the phase difference in the interferometer. The storage time was set to 1.6 $\mu$s. For this particular fringe, a visibility of V=91.5$\%$ is reached, limited by phase noise caused by vibrations in the cooling system.}
\label{figure_9}
\end{figure}
\end{center}
\ewt

\begin{figure}[b]
\centering
 \includegraphics[width=.4\textwidth]{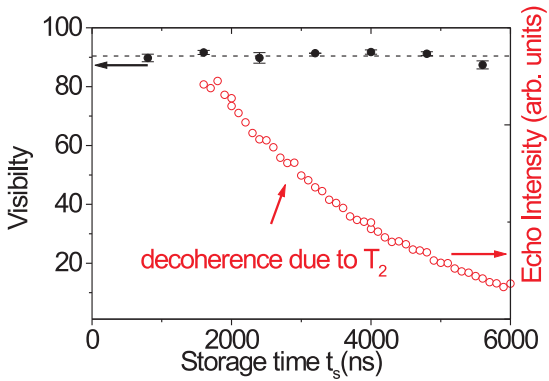}
 \caption{Interference fringe visibility (filled circles) is shown as a function of the memory storage time. Atomic decoherence strongly acts on the amplitude of the echo signal, as shown here for one of the waveguides  (open circles), but leaves the visibility unaffected. The dotted line shows the average visibility of 90.5$\%$.}
\label{figure_10}
\end{figure}

An important aspect of both experiments reviewed above is the
robustness of the phase coherence of the optical storage with
respect to the atomic decoherence processes, for instance magnetic spin interactions between the RE ions (see section \ref{Material considerations}). The loss of atomic coherence is generally characterized by the atomic coherence time $T_2$. Intuitively, one might expect that the phase coherence of optical pulses retrieved from the memories would be reduced if the storage time was of the order of or longer than $T_2$. In both experiments, however, the interference visibility turned out to be independent of the storage time. In the second experiment \cite{Staudt2007b}, the visibility of the interference was explicitly investigated as a function of storage time, see Fig. 10. Although the efficiency of the storage was significantly reduced when the storage time was of the order of $T_2$, the visibility remained at constant value of $\sim90\%$. This property can be attributed to a collective enhancement effect where the photon-echo emission is given by the sum of emission amplitudes from atoms that have retained their coherence. We separate the total number of atoms $N$ into an incoherent sub-ensemble of atoms $N_{inc}$ that have interacted with the environment  and a coherent sub-ensemble of atoms $N_{coh}$ that are still phase coherent. As time evolves, the size of the coherent sub-ensemble well decreases, while the size of incoherent sub-ensemble will increase with a time constant of $T_2$. The sum expressing the coherent emission probability in the spatial mode of the photon-echo runs over all atoms that are in phase $N_{coh}$ and it therefore scales as $(N_{coh})^2$. Emission from the incoherent atoms into the same spatial mode, however, scales as $N_{inc}$, such that the signal-to-noise ratio is proportional to $N_{coh}^2/N_{inc}$, which, for solid state ensembles of RE ions is typically very large ($N > 10^6$). The collective enhancement effect thus ensures that the echo mainly stems from the still coherent sub-ensemble of atoms, i.e. atoms that have undergone no or little phase perturbations. Collective enhancement is an underlying principle in most quantum memory proposals implemented in large ensembles of atoms, which is also the case for CRIB memories in RE doped solids. The independence of qubit fidelity with respect to atomic phase relaxation has also been derived using the Schr\"{o}dinger equation, see \cite{Moiseev2008b}.


\section{Experimental realizations of CRIB}
\label{experimental CRIB}

The main issues to be addressed when considering demonstrating a CRIB based optical memory is how to induce a controlled and reversible inhomogeneous broadening on an optical transition, and that the broadened transition must have sufficient optical depth to absorb a significant component of the input pulse. In this section we review how these conditions have been met using the linear Stark effect in RE doped crystals.

Currently available samples of rare-earth doped crystals that have a large linear Stark shift suitable for CRIB also have inhomogeneous linewidths of the order of a GHz (or more). The mechanisms contributing to this broadening have been discussed in section \ref{Material considerations}. Given that this linewidth is comparable to the maximum Stark induced broadening possible, and would also necessitate the switching of the electric field gradient on a sub-nanosecond time scale, it has not been possible to demonstrate CRIB without first modifying the inhomogeneous line profile of the transition to create a narrow spectral feature through optical pumping \cite{Pryde2000,Nilsson2004,Crozatier2004,Rippe2005,Wesenberg2007,Alexander2006,Hetet2008,Hastings2008,Rippe2008}. Until samples become available with much narrower inhomogeneous linewidths it is likely that demonstrations of CRIB in solid state systems will be restricted to transitions that exhibit efficient and long lived spectral holeburning. To date CRIB based memories have been reported in two solid state systems: on the $^{7}$F$_{0}$$\leftrightarrow$ $^{5}$D$_{0}$ transition in Eu$^{3+}$:\YSO \cite{Alexander2006,Alexander2007}, and on the $^{3}$H$_{4}$$\leftrightarrow$$^{1}$D$_{2}$ in Pr$^{3+}$:\YSO \cite{Hetet2008}. For both these transitions the holeburning mechanism involves optically pumping population into long lived ground state hyperfine levels. A consequence of using this spectral holeburning mechanism, to create the required narrow feature, is that the operation of the memory is limited to a bandwidth less than the hyperfine splittings, which for both crystals is of the order of 10 MHz. The use of Tm$^{3+}$:Y$_{2}$O$_{3}$, where the holeburning involves a metastable electronic state, has been proposed to avoid this problem \cite{Louchet2007}. Another alternative would be to use the Zeeman level structure of odd-electron RE ions such as Neodymium \cite{Hastings2008,Staudt2008} and Erbium \cite{Hastings2008a}, where level spacings larger than 1 GHz can be obtained by applying moderate magnetic fields.

\bwt
\begin{center}
\begin{figure}[t]
 \includegraphics[width=0.75\textwidth]{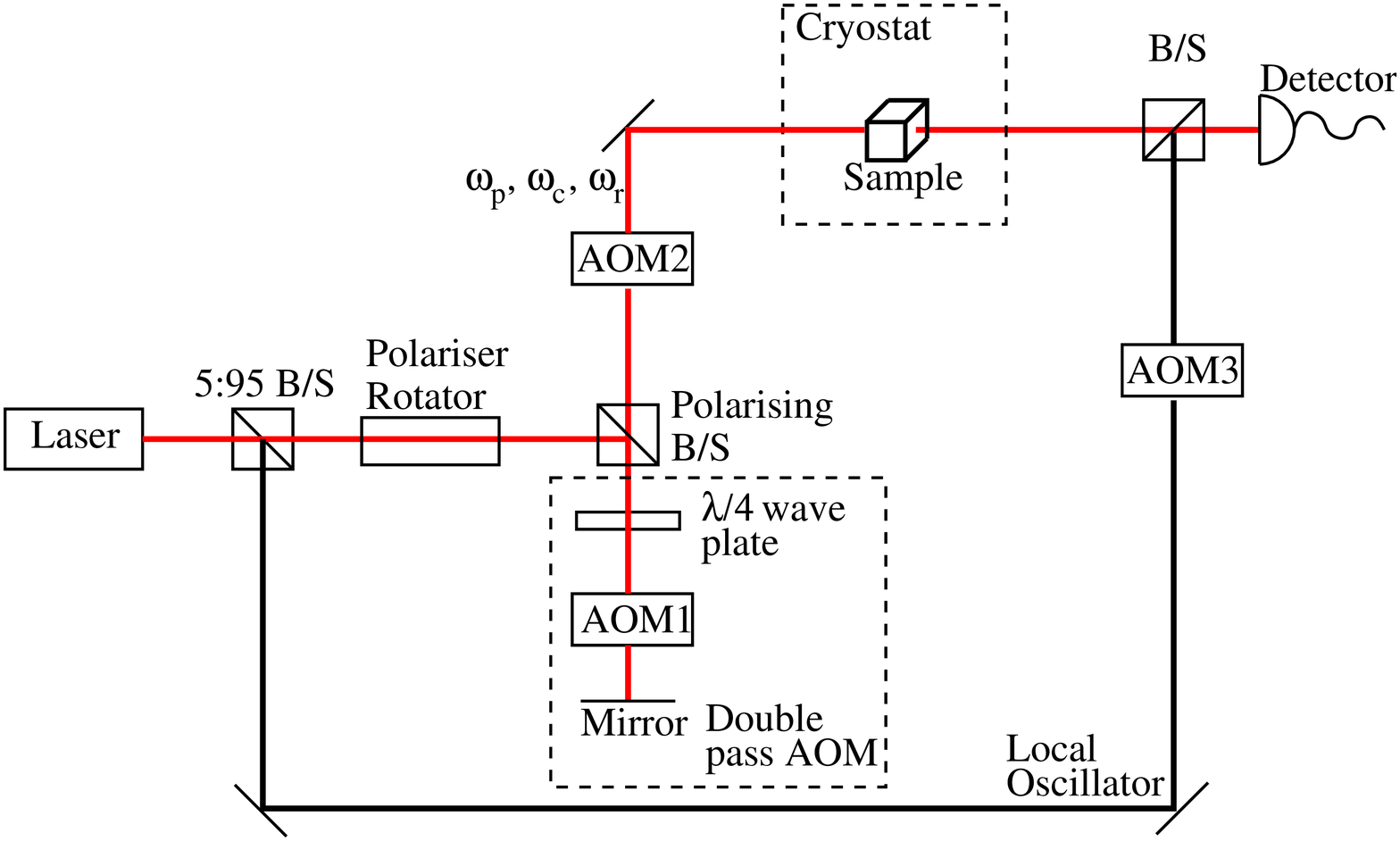}
 \caption{Experimental set-up for the two-level gradient Stark echoes. Beam splitters are labeled as B/S and acousto-optic modulators are labeled as AOM.}
\label{figure_11}
\end{figure}
\end{center}
\ewt

\begin{figure}[!ht]
 \includegraphics[width=0.35\textwidth]{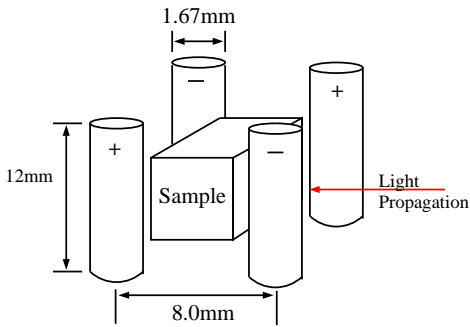}
 \caption{Experimental set-up of the electrodes and the sample. This arrangement of the electrodes produced a quadrupolar electric field across the sample with an electric field gradient of 225~Vcm$^{-2}$ when $\pm$35~V was applied to the electrodes.}
\label{figure_12}
\end{figure}

Some early proposals for CRIB based memories have the required controlled inhomogeneous broadening induced on a material level, i.e. utilize a microscopic variation in how the ions respond to the applied electric field. An example how the latter could be achieved was suggested by Hastings et al. \cite{Hastings-Simon2006}, and exploits specific (lack of) symmetry in RE glasses, leading to transverse inhomogeneous broadening under application of a constant constant electric field (see section \ref{Material considerations}). An alternative approach to induce the controlled broadening is to apply a field gradient across a crystal, as proposed in \cite{Nilsson2005a,Kraus2006}. Depending on the direction of the gradient, transverse as well as longitudinal broadening can be achieved. All of the demonstrations of CRIB performed to date have relied on the latter method, i.e. longitudinal broadening has been induced through the application of a macroscopic field gradient. A practical difference between transverse and longitudinal broadening is that for the former efficient echoes can only be obtained for echoes propagating in the backwards direction whereas for the later efficient echoes can be obtained in the forwards direction as long as the electric field varies monotonically across the sample (see Eqs. \ref{recall efficiency transverse backwards}, \ref{recall efficiency transverse forwards} and \ref{recall efficiency longitudinal}). This specific realization of CRIB is often referred to as gradient echo memory (GEM).

Fig. 11 shows the experimental setup used by Hedges and co-workers \cite{Hedges2008} to demonstrate a gradient echo memory with forward recall efficiencies up to 45\% in 0.05 at\% $^{141}$Pr$^{3+}$:\YSO. The optical transition $^{3}$H$_{4}$$\leftrightarrow$$^{1}$D$_{2}$ at 605.977~nm was excited with linearly polarized light propagating along the C$_{2}$ axis of the crystal, with the polarization chosen to maximize the absorption, which was 140 dB/cm at the centre of the natural, inhomogeneous line.  The length of the crystal in the direction of propagation was 4~mm.  The crystal was cooled to below 4~K in a liquid helium bath cryostat. The frequency and intensity of the light incident on the sample was controlled with two acousto-optic modulators (AOMs) in series. A Mach-Zehnder interferometer arrangement with the AOMs and sample in one arm was employed to enable phase detection of the coherent emission from the sample, when required. A linear electric field gradient in the light propagation direction was applied to the sample using four 12~mm long, 1.7~mm diameter rods in a quadrupolar arrangement, as shown in Fig. 12. A narrow spectral feature was produced using an optical pumping procedure that consisted of burning a relatively wide ($\approx$4~MHz) spectral hole in the absorption line by scanning the laser frequency.  A 200kHz wide antihole was placed in the middle of this region by applying optical excitation at frequency offsets at $\pm$10.2~MHz, $\pm$27.5~MHz and $\pm$17.3~MHz, corresponding to the groundstate hypefine splittings.

Shown in Fig. 13 is a two-level gradient echo. A 1.1~$\mu$s optical input pulse excited a 200~kHz feature that had been broadened to 1~MHz through the application of $\pm$35~V to the electrodes.  The polarity of the electrodes was reversed after a delay. The optical depth of the broadened transition was 0.8. The decay of the echo with increasing delay reflects the residual 200 kHz linewidth of the feature in zero electric field.

\begin{figure}[!ht]
\includegraphics[width=0.45\textwidth]{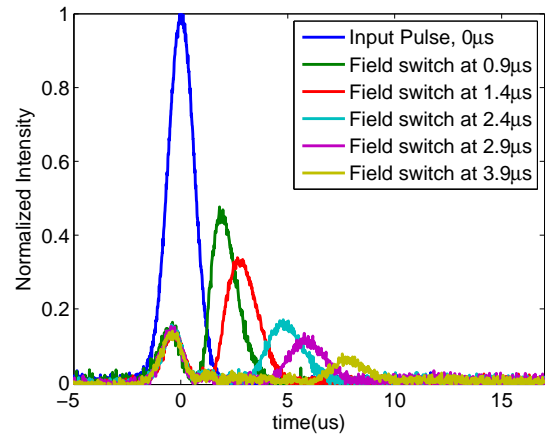}
 \caption{Transmitted and retrieved pulses stored in identically prepared atomic ensembles with increasing delay times: 1.5~$\mu$s, 2.5~$\mu$s, 3~$\mu$s and 4~$\mu$s. The time axis for the traces are adjusted so that the input pulse occurs at 0. Also shown is the unabsorbed pulse passing through an empty spectral region of the crystal. The maximum efficiency is 45\%.}
\label{figure_13}
\end{figure}

\section{Conclusion and outlook}
\label{conclusion}

The development of quantum memory for quantum repeaters requires experimental and theoretical skills that stretch across a variety of intellectual and technical borders, including quantum communication, quantum optics, and materials science, making it a very interesting and challenging field to work in. The photon-echo quantum memory, or CRIB, employs an atomic ensemble with controllable inhomogeneous broadening, and is thus naturally in the direct lineage of traditional photon-echo and coherent transient experiments that have been perform for decades. The possibility to build on this heritage greatly benefits the development of CRIB, in particular for early tests or in the quest to tailor, find, or create, the appropriate solid state material, possibly RE ion doped crystals.

A photon-echo quantum memory represents an interesting alternative to a single atomic system approach \cite{Cirac1997}. Obviously, using quantum state transfer between a photon and an ensemble of inhomogeneously broadened atomic absorbers increases the complexity of the system, and has required, and will continue to require, the development of new theoretical approaches and novel experimental tools. This article reviews the historical development from simple photon-echoes to the latest CRIB based storage of classical optical pulses, and puts photon-echo quantum memory into the context of quantum repeaters. Clearly, the complexity of the system can be turned into richness in order to be controlled and used for storage or quantum state processing purposes. Indeed, the manipulation of the inhomogeneous broadening during the photon absorption and re-emission stage is only limited by our imagination and our ability to control the medium. Protocols that go beyond simple time reversal of the absorption process \cite{Staudt2007a,Moiseev2008b,Delfan2008,Underwood2008}, combine controlled reversible inhomogeneous broadening with other storage approaches \cite{Hetet2008b,Moiseev2008c}, and making better use of the available atomic density for storage of multi-mode optical fields \cite{Afzelius2008b} are currently being developed. It is foreseeable that the high degree of quantum control over the atom-light evolution and the high efficiency of these new protocols may not only allow future quantum information processing, but may also benefit classical data storage and manipulation.

\section*{Acknowledgements}
The idea to write this review article in an international collaboration featuring researchers from seven different groups and countries, and representing all possible aspects of photon-echo quantum memory, arose during a workshop in Bozeman, Montana, USA in January 2008. The article is meant to represent the entire research groups, and the authors gratefully acknowledge the support of all colleagues without whom this article would never have been possible.

The authors gratefully acknowledge financial support by the Natural Sciences and Engineering Research Council of Canada (NSERC), General Dynamics Canada, Alberta's Informatics Circle of Research Excellence (iCORE), the Swiss NCCR Quantum Photonics, the European Commission through the integrated project QAP, the U. S. Air Force Research Laboratory (US Air Force Office of Scientific Research), the U. S. Army Research Office, the Montana Board of Research and Commercialization Technology, the Royal Swedish Academy of Science, the Swedish Research Council, and the Russian Foundation for Basic Research (grant no. 06021682).

\def\bstname{lpr}

\end{document}